# Subpicotesla Scalar Atomic Magnetometer with a Microfabricated Cell

Rui Zhang,[1,a)] Terry Dyer,[2)] Nathan Brockie,[3)] Roozbeh Parsa,[4)] and Rahul Mhaskar[1)]

[1]*Geometrics, Inc., San Jose, CA, 95131, USA*

[2] *formerly Texas Instruments, Greenock, PA16 0EQ, UK, now University of Strathclyde, Glasgow, G11 XQ, UK*

[3] *formerly Texas Instruments, Greenock, PA16 0EQ, UK, now Nu nano Ltd, Bristol, BS2 0XJ, UK*

[4] *formerly Texas Instruments, Santa Clara, CA, 95051,USA, now Rockley Photonics Inc, San Jose, CA,95110,  USA*

We demonstrated a scalar atomic magnetometer using a micro-fabricated Cs vapor cell. The atomic spin precession is driven by an amplitude-modulated circularly-polarized pump laser resonant on D1 transition of Cs atoms and detected by an off-resonant linearly-polarized probe laser using a balanced polarimeter setup. Under a magnetic field with amplitude in the Earth's magnetic field range, the magnetometer in the gradiometer mode can reach sensitivities below 150 fT/√Hz, which shows that the magnetometer by itself can achieve sub-100 fT/√Hz sensitivities. In addition to its high sensitivity, the magnetometer has a bandwidth close to 1 kHz due to the broad magnetic resonance inside the small vapor cell. Our experiment suggests the feasibility of a portable, low-power and high-performance magnetometer which can be operated in the Earth's magnetic field. Such a device will greatly reduce the restrictions on the operating environment and expand the range of applications for atomic magnetometers, such as detection of nuclear magnetic resonance (NMR) in low magnetic fields.

## I.    INTRODUCTION

Atomic sensors have been well known for their ultra-precise measurement of many important physical quantities[1], such as time, angular velocity and magnetic field. However, they have not been widely used yet in commercial applications mostly due to their high cost, large size and high power consumption. In recent years much progress has been made towards miniaturization of atomic sensors[2], especially in atomic clocks[3]. The first chip scale atomic clock became commercially available in 2011. Chip scale atomic magnetometers, constructed using the technique of Micro-Electro-Mechanical Systems (MEMS), have also been demonstrated in laboratories[4]. MEMS scalar atomic magnetometers can reach sensitivities of 5pT/√Hz. More recently, using a MEMS-fabricated vapor cell, researchers achieved sensitivity below 100 fT/√Hz with an atomic magnetometer operating in the spin-exchange-relaxation-free (SERF) regime[5]. Advances in MEMS based research can eventually lead to small, low-cost, and power efficient atomic magnetometers since MEMS technology not only produces much smaller devices but also makes mass production possible at reduced costs. Here we demonstrate a scalar atomic magnetometer using a MEMS-based Cs vapor cell. The achieved sensitivity is comparable to the SERF MEMS magnetometer with the advantage of being operational in much larger and wider range of magnetic fields. Similar results were recently demonstrated in a slightly smaller Rb vapor cell with about 300 fT/√Hz sensitivity[6] and using much bigger cells scalar magnetometers with better than 20 fT/√Hz sensitivity have been achieved[7]. Comparable sensitivity performance is also expected for a recently proposed atomic magnetometer system[8].

a)    Email: rzhang@geometrics.com

## II. EXPERIMENTAL SETUP

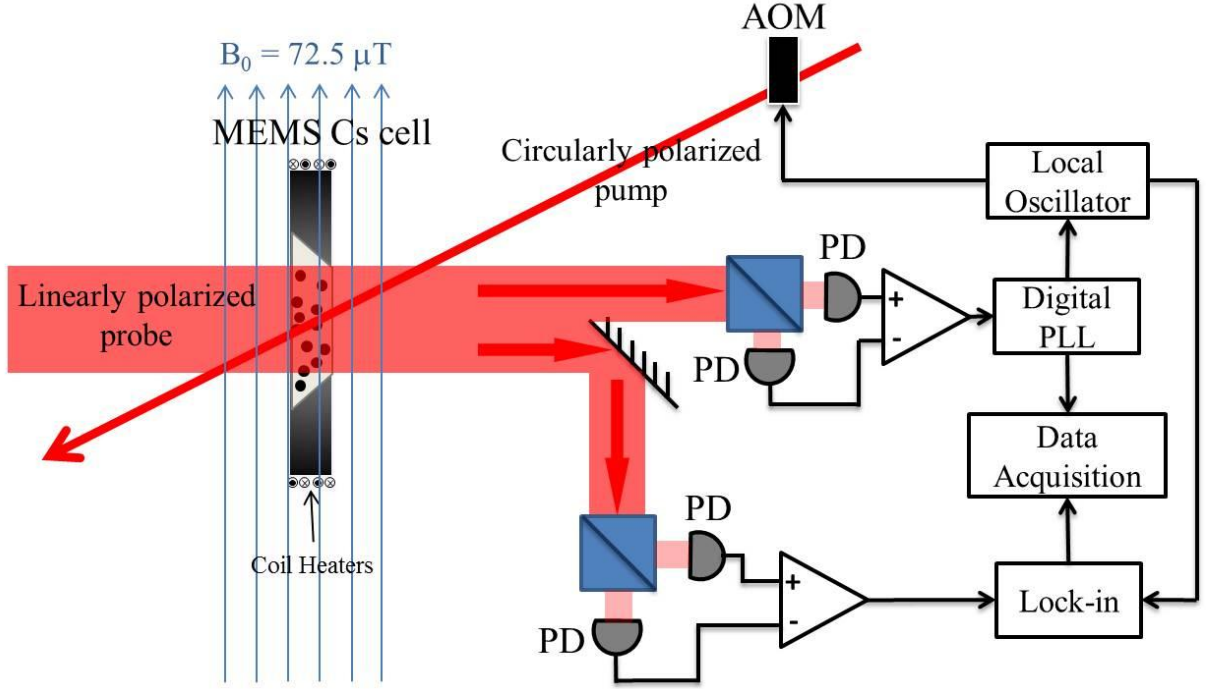

Figure 1 Schematics of the experimental setup for gradiometer measurement using a MEMS cell. The cell has an inner dimension of 4.5mm x 3mm x 1.5mm. The probe and pump beams are counter-propagating and cover the 4.5mm x 3mm window of the cell. After the cell, the probe beam is split into two parts for gradiometer measurement.

Our experimental setup is shown schematically in Figure 1. Pump and probe beams are originated from two distributed Bragg reflector (DFB) lasers. The pump light is circularly polarized and passed through an acousto-optical modulator (AOM) before overlapped with the linearly polarized probe beam at the Cs vapor cell, which is located inside a 4-layer magnetic shield can (not shown). The Cs vapor cell was manufactured by Texas Instruments using MEMS technology. The cell has a trapezoidal structure with minimum inner dimensions 4.5mm x 3mm x 1.5mm, the shortest distance being in the light path direction. The cell is also filled with about 50 torr nitrogen gas as the standard buffer gas to reduce the collision between the Cs atoms and the walls. The cell is heated to about 110°C by running an AC current (10 kHz) through external coils. The coils are wound such that the AC current generates minimal stray magnetic field. The pump and probe beams are counter-propagating and have a Gaussian beam waist of 3 mm and 4 mm in diameter, respectively (In Figure 1, beam sizes are not shown proportionally). Pump light has 1mW in total power before the cell with AOM continuously on (0.6 mW after the cell at room temperature) and its frequency is tuned to the $|F=3\rangle \rightarrow |F'=4\rangle$ transition of Cs D1 line. When the pump light is on, a spin polarization in the $|F=4\rangle$ state is generated. In the presence of a magnetic field perpendicular to the light, the atomic ground state spin undergoes Larmor precession. When the frequency of the periodic pumping, controlled by the amplitude modulation (AM) of the AOM, is synchronized with the Larmor precession frequency, the spin population has a resonance, called the magnetic resonance. The magnetic resonance can be detected by the probe light. The probe light is linearly polarized, 0.7 mW in total power before the cell and about 4 GHz blue detuned with respect to the |F=4> → |F'=3> transition of Cs D1 line. After passing through the Cs cell, the probe light power is reduced to about 210 µW and is spatially split into two beams, each entering a standard balanced polarimeter setup for detection of the Larmor precession. The differential photodiode currents are then amplified through transimpedance amplifiers with a gain of 100 kΩ. Out of the amplifier, one of the Larmor precession signals is fed into a lock-in amplifier and the other into a digital phase lock loop (PLL). When the PLL is enabled and its local oscillator signal is fed into the reference input of the lock-in amplifier, the output of the lock-in amplifier becomes the differential reading between the two Larmor precession signals, thus the gradiometer measurement. A similar setup was previously discussed[9].

## III. EXPERIMENTAL RESULTS

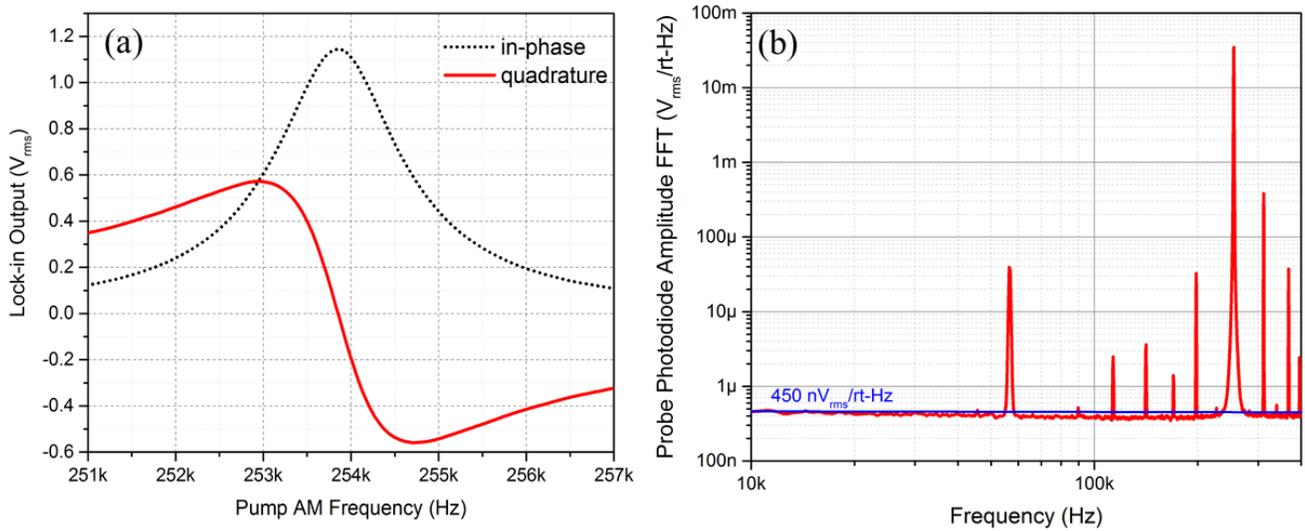

Figure 2 (a) Probe lock-in signal (r.m.s. amplitude) as a function of pump AM frequency $f$. The black dot (red solid) curve represents the in-phase (quadrature) component. (b) Amplitude spectral density of the probe photodiode signal after the transimpedance amplifier.

When the PLL is not enabled, the pump AM frequency can be scanned around the Larmor frequency and the outputs of the lock-in amplifier can be recorded as a function of the pump AM frequency. In the presence of a perpendicular magnetic field of $B_0 = 72.5 \mu T$, generated by a pair of Helmholtz coils inside the shield can, the observed magnetic resonance is shown in Figure 2 (a). The phase of the lock-in amplifier is adjusted such that the quadrature component is zero (has the maximum slope) at the resonance. The resonance has a FWHM of about 1.8 kHz and the slope of the quadrature component is about 1.35mV/Hz at the resonance. To measure the magnetic field in the open loop mode, the pump AM frequency is set at the center frequency, $f_0$, of the magnetic resonance. Then the quadrature output of the lock-in amplifier can be converted to the magnetic field B according to the following equation: $B = (f_0 + V/L)/k$, where V is the quadrature output, L=1.35mV/Hz is the slope and k=3.5 Hz/nT is the Zeeman splitting factor between two adjacent magnetic sublevels of Cs atoms in |F=4> ground state. Similarly the magnetometer output noise can be measured in the open mode as $\delta B = (\delta V/L)/k$, where $\delta V$ is the quadrature output noise. Amplitude spectral density of the amplified probe photodiode signal is plotted in Figure 2 (b), the noise floor of which is a good estimation of $\delta V$. Based on the 450 nV/√Hz noise floor shown in Figure 2 (b), the magnetometer output noise should be around 95 fT/√Hz. However, here we assume that $f_0$ is constant, which is often not the case especially for a magnetic field as strong as the one we are using. The current noise of the electronics that generates the $B_0$ is often the dominating noise. Therefore, to measure the fundamental noise of the magnetometer, we need to use the gradiometer setup as shown in Figure 1.

To operate the system in the gradiometer mode, the PLL is enabled to track the magnetic field $B_0$. The PLL output, $f_{PLL} = f_0+\Delta f$, contains both the fluctuation of the $B_0$ and a slight frequency offset due to the magnetic field gradient. When $f_{PLL}$ is fed into the reference input of the lock-in amplifier, the common fluctuation in $f_0$ is cancelled, leaving behind only the fundamental noise of the magnetic field gradient. The fundamental noise of the gradient is a factor of √2 larger than that of an individual magnetometer, assuming the same fundamental noise for both paths. In the gradiometer experiment, we record both the PLL output (individual magnetometer reading) and the lock-in amplifier output (magnetic field gradient) and compare their noise levels, shown in Figure 3 as the black curve and the cyan curve (overlapping with the red curve at low frequencies), respectively. As we can see, the gradiometer output has a noise level as low as 150 fT/√Hz, much lower than the individual magnetometer measurement. The fundamental noise of each individual magnetometer should be around 105 fT/√Hz based on the gradiometer result, which is very close to the 95 fT/√Hz estimation based on the noise floor of Figure 2 (b). Photon shot noise is often the most dominant noise source of the atomic magnetometer. For a light power of $p$, the photon shot noise has a flat amplitude spectral density of $\sqrt{(2h\nu p)}$, where $h\nu$ is the single photon energy. With $p$ = 105 μW, λ = 895 nm, photodiode efficiency of 0.65 A/W and 100 kΩ transimpedance amplifier gain, the noise floor of the amplitude spectral density due to the photon shot noise is 446 nV/√Hz, leading to a magnetometer noise of about 94 fT/√Hz. Therefore the fundamental scalar magnetometer noise is indeed dominated by the photon shot noise.

The probe beam is split into two for the gradiometer measurement. If the full probe beam is used, there should be another factor of √2 improvement in the fundamental scalar magnetometer noise, provided that the noise of the individual magnetometer shown in Figure 3 is dominated by fluctuation in the background magnetic field. Based on the noise floor shown in Figure 2 (b), the individual scalar magnetometer noise is not due to the noise in the probe or pump beams, leaving the background magnetic field fluctuation the most likely cause. With a better current source for the background magnetic field, we are able to achieve better than 180 fT/√Hz noise level for the individual scalar magnetometer. Therefore we are confident that we should be able to achieve better than 100 fT/√Hz noise performance with the micro-fabricated cell. The increase of the noise level below 30 Hz shown in Figure 3 is mainly due to air current along the beam paths, causing fluctuation of the air refractive index, which affects the polarization of the beams. This negative effect can potentially be eliminated using vacuum packaging technique. The sharp spikes in Figure 3 are due to the couplings of the 60 Hz power line radiation and its harmonics into the electronics.

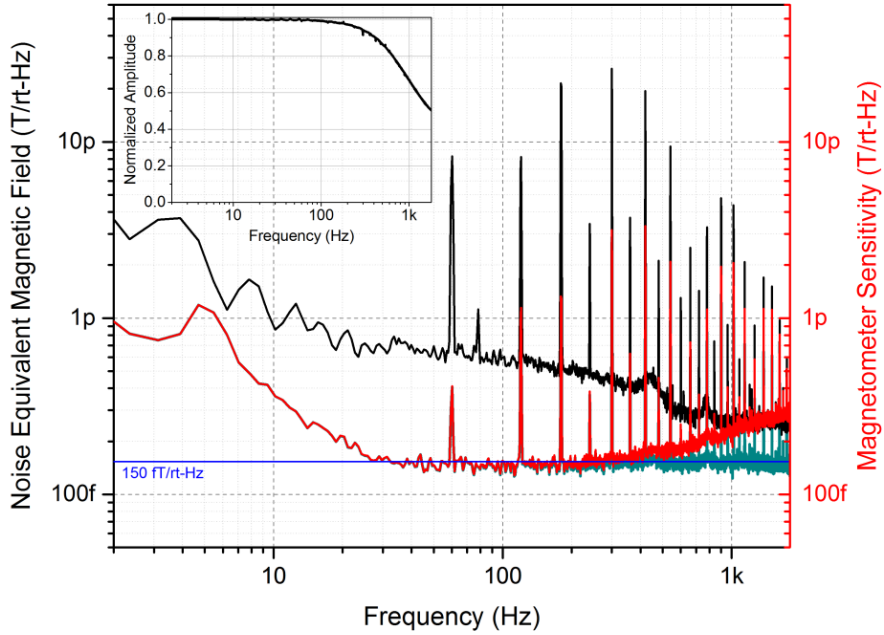

Figure 3 Spectral density of the magnetometer noise in the gradiometer mode (Cyan curve, overlapping with the red curve at low frequencies). Inset shows the bandwidth of the magnetometer in the open loop mode. The red curve is the fundamental sensitivity of the magnetometer, which represents the minimum signal level detectable by the magnetometer with a signal to noise ratio of 1:1. The black curve is the noise spectral density of the individual magnetometer operating in the phase-lock-loop mode.

Another important performance characteristic of a magnetometer is its bandwidth. To measure the bandwidth, we have another set of Helmholtz coil in the same direction as $B_0$, which can independently generate an oscillating magnetic field, $B_1 \sin(2\pi f_1 t)$. We fix the amplitude $B_1$ at about 3nT and vary the frequency $f_1$. The oscillating magnetic field can be detected by the magnetometer in open mode and the measured amplitude $B_1$ as a function of $f_1$ is recorded. The normalized $B_1$ vs $f_1$ is plotted in Figure 3 inset. As shown, the 3dB-point of the bandwidth curve is about 900 Hz. The bandwidth is limited by the width of the magnetic resonance shown in Figure 2. By combining the magnetometer noise spectrum and its bandwidth, we can achieve the sensitivity of the magnetometer. We define the sensitivity as the minimum detectable signal level by the magnetometer at a signal to noise ratio of 1:1. According to this definition, the sensitivity curve can be calculated by dividing the noise spectrum by the bandwidth curve, which is shown as the red curve in Figure 3. At low frequencies, the sensitivity and the noise curves overlap. At higher frequencies where the signal starts to drop due to the bandwidth, the sensitivity is getting worse since the noise remains almost the same. Based on the sensitivity curve, for a 300 fT/√Hz performance, the gradiometer can be operated up to 2 kHz.

## IV.　DISCUSSION AND CONCLUSION

The combination of high sensitivity, large bandwidth and operation in the Earth's magnetic field, in addition to the possibility of small size and low power consumption associated with the micro-fabricated cell, makes the demonstrated magnetometer attractive to many applications[10]. For example, using the atomic magnetometer for detection of nuclear magnetic resonance (NMR) in low magnetic fields[11] is becoming an interesting research topic due to many advantages brought forth by the low operational magnetic field and the compact size of the atomic magnetometers. However, one of the major challenges arises due to the several orders of difference between the gyromagnetic ratios of electrons and nucleus[12]. The NMR signal often falls far outside the bandwidth of the atomic magnetometer if the NMR sample and the magnetometer are in the same magnetic field. People have used smart ways to circumvent this difficulty[11], but at the cost of increased complexity of the setup. A highly sensitive scalar magnetometer with near 1 kHz bandwidth and using a micro-fabricated cell can potentially solve many problems in the NMR application. First, due to the high bandwidth, the magnetometer and the NMR nucleus can be operated in the same magnetic field. Taking one of the commonly used nucleus in NMR, $^{129}$Xe, for example, at a magnetic field of 50 μT, the nucleus precession has a frequency of about 600 Hz, well within the bandwidth of the scalar magnetometer. Second, the nucleus can be brought much closer to the magnetometer since now they can be in the same magnetic field. Combined with the micro-fabricated cell and the micro-fluid channels[13], it is possible to reduce the separation between the cell and nucleus to within 1 mm, greatly enhancing the NMR signal. In addition, the scalar magnetometer can be operated in a wide range of magnetic fields, offering tuning of the NMR frequency if necessary. Depending on the strength of the NMR signal, NMR detection may become possible in an unshielded environment using the magnetometer in the gradiometer mode. The demonstrated scalar atomic magnetometer may well lead to a portable and low-cost NMR device in the near future.